\documentclass[aip,rsi,reprint]{revtex4-1}
\usepackage{graphicx}
\usepackage{amsmath}
\usepackage{amsfonts}
\usepackage{amssymb}
\usepackage{graphicx}
\usepackage{dcolumn}
\usepackage{bm}

\begin{document}

\title{Microstrip filters for measurement and control of superconducting qubits}

\author{Luigi Longobardi}
\affiliation{American Physical Society, 1 Research Road, Ridge, New York 11961, USA}
\affiliation{Department of Physics and Astronomy, Stony Brook University, Stony Brook, New York 11794--3800}
\email{llongobardi@ms.cc.sunysb.edu}
\author{Douglas A. Bennett}
\affiliation{National Institute of Standards and Technology, Boulder, Colorado 80305, USA}
\affiliation{Department of Physics and Astronomy, Stony Brook University, Stony Brook, New York 11794--3800}

\author{Vijay Patel}
\affiliation{Department of Physics and Astronomy, Stony Brook University, Stony Brook, New York 11794--3800}

\author{Wei Chen}
\affiliation{Department of Micro/Nanoelectronics, Tsinghua University, Beijing, 100084, People's Republic of China}

\author{James E. Lukens}
\affiliation{Department of Physics and Astronomy, Stony Brook University, Stony Brook, New York 11794--3800}

\date{\today}

\begin{abstract}
Careful filtering is necessary for observations of quantum phenomena in
superconducting circuits at low temperatures.
Measurements of coherence between quantum states requires
extensive filtering to protect against noise coupled from room
temperature electronics. We demonstrate distributed transmission line filters which cut off
exponentially at GHz frequencies and can be anchored at the base
temperature of a dilution refrigerator. The compact design makes them suitable to
filter many different bias lines in the same setup, necessary for
the control and measurement of superconducting qubits.

\end{abstract}

\maketitle

\section{introduction}

Josephson systems have proven to be a fantastic test-bench for the study of fundamental quantum phenomena and for the realization of circuits for quantum information processing\cite{clarke2008}.
 In order to observe quantum effects such as quantum tunneling\cite{devoret1985}, phase diffusion\cite{kivioja2005}, quantum coherence\cite{bennett2009} and Rabi oscillations\cite{Martinis2002}, the device must be well isolated from the environment and all measurement and control lines must be heavily filtered to reduce the level of noise from room temperature electronics. The fragile nature of quantum coherence in superconducting qubits
places strict requirements on the design of the device, its readout
and the experimental apparatus. A common configuration for the bias and readout lines is to have a sequence of filters thermally anchored at different temperature stages of a dilution refrigerator \cite{tian:2012}. For instance, in our quantum coherence experiment \cite{bennett2009}, the low frequency lines were filtered by electromagnetic interference (EMI) filters at room temperature, followed by cascaded RC filters anchored at the 1.4K stage, and finally by specially designed microstrip filters at base temperature, while for high frequency lines we used a series of attenuators at 1.4K, 600 mK and at base temperature followed by microstrip filters that cutoff around 1 GHz.

Common RC filters only work up to a resonant frequency due to the parasitic inductance of the
capacitor \cite{cain}. The parasitic inductance of the capacitor is
modeled as a lumped inductance in series with the capacitance. In an
RC circuit the capacitor shorts high frequency signals to ground.
However when the parasitic inductance is included the high frequency
signal is blocked from traveling through the series combination. The
frequency at which the impedance due to the inductance equals
the impedance due to the capacitance is given by
\begin{equation} f_{res} = \frac{1}{2 \pi \sqrt{L_{para} C}} \end{equation}
High quality high-frequency chip capacitors usually have a parasitic
inductance of around 500 pH \cite{murata,cain}. A filter realized with a 1 nF capacitor has a resonant frequency around 225 MHz,  making it unfit to filter high frequency noise in the GHz range and above.

Transmission line filters are used to avoid this problem with
parasitic inductance in RC filters. A lossy transmission line can be
modeled as series inductance (L) and resistance (R) per unit length
and shunt capacitance (C) and conductance (G) per unit length. The
voltage transmitted through the transmission line is given by\cite{ramo}
\begin{equation} V = V_0 e^{- 2 \alpha z}  \label{transmitted} \end{equation}
where $V_0$ is the initial voltage, z is the propagation distance,
and $\alpha = \Re(\gamma)$ is the attenuation constant and the
propagation constant ($\gamma$) is defined as
\begin{equation} \gamma = \sqrt{(R + j \omega L)(G + j \omega C)}  \label{propconst} \end{equation}
with $\omega$ the frequency of the signal. For the cases
of our interest it can be assumed that $G=0$.

At high frequencies,
$R \ll \omega L$. Using the binomial expansion we obtain
\begin{equation} \alpha = \frac{1}{2} \frac{R}{Z_0} \end{equation}
where $Z_0 =\sqrt{L/C}$ is the characteristic impedance.  In the frequency range when $\omega L \ll R$ the propagation constant becomes $\gamma = \sqrt{j \omega R}$. By considering the real part we obtain
\begin{equation} \alpha = \frac{\sqrt{2 \omega R C}}{2} \end{equation}
Under the condition $\alpha *z \ll 1$, the frequency dependent term in Equ. \ref{transmitted} is approximately constant and the transmission line can be treated as a lumped resistance.
For a transmission line of length $\ell$ this condition can be rewritten as  $\omega \ll \frac{2}{\ell^2} \frac{1}{R C}$.

There are a number of distributed filters that can operate in this
frequency range including copper powder filters \cite{martinis1987,bladh,milliken:024701,lukashenko:014701}, thermocoax\cite{zorin1995} and
microstrip filters \cite{vion1995,sueur:115102,courtois:1995}.  Copper powder filters work by
using the skin effect in the individual copper grains to attenuate
high frequencies \cite{fukushima}. Thermocoax uses the
 resistance of the cable to form a lossy transmission line\cite{zorin1995}. In this paper we report on the design, realization, and testing of lossy microstrip filters with a faster roll off compared to copper powder filters and that are more compact than thermocoax lines. Our microstrip filters were designed to be compact to fit within the space constraints typical of a dilution refrigerator  \cite{mandal2011}, attenuate strongly at high frequencies and constructed to be microwave tight.

\section{Design and Fabrication}

Our microstrip filters consist of a thin film of chromium deposited on a substrate and
diced into long narrow chips.  When the chips are placed in a brass
housing, which acts as a ground plane, they form a lossy microstrip
transmission line. Some of the bias currents for our qubit require
currents of few milliamperes. At these currents any non-negligible
resistance could cause heating of the filters and the sample \cite{bluhm:2008}. To
avoid heating in the filters that need to pass dc currents, a
superconducting shunt is deposited on top of the chromium
microstrip. The shunt is patterned as a meander line to provide an
inductance large enough to block high frequency signals while
allowing dc currents to pass through without resistance. At
frequencies above the cutoff off of the meander line, the current travels through the chromium microstrip and is filtered.

The flexibility of our microstrip design allows for the realization of impedance-matched filters \cite{Santavicca2008,Slichter}. The characteristics of this lossy transmission line can be estimated
from the structure of the microstrip. The impedance $Z_0$ of the microstrip is dominated by its width (w), the height of the strip above the ground plane (h) and the
dielectric constant of the substrate material ($\epsilon_r$). Approximate values of $Z_0$ can be found in the literature \cite{ramo,gupta}.
\begin{figure}
\centering
\includegraphics[width=3.3in]{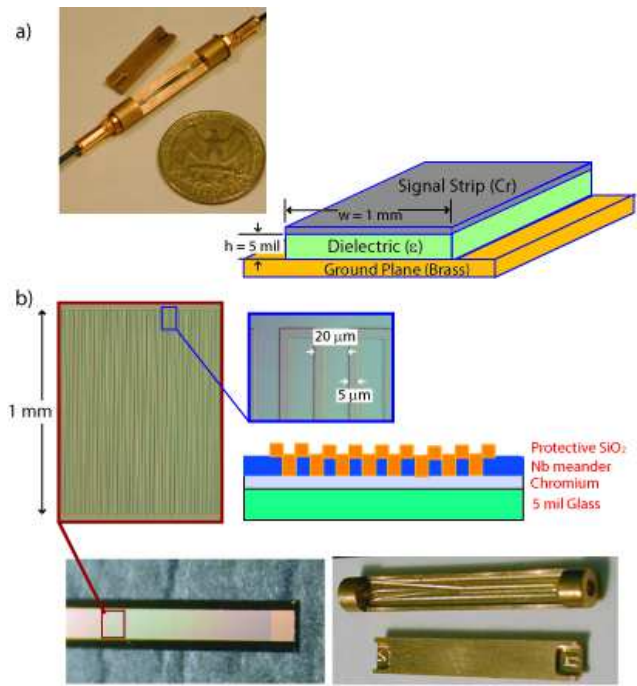}
\caption{\label{microstrip} (Color Online) Pictures and sketches of the lossy microstrip filter
a) without and b) with the Nb meander line, and their respective
housings.}
\end{figure}

A cross section of the microstrip filters is shown in Fig. \ref{microstrip}.
The fabrication process is rather straightforward.
The chromium is evaporated on 5 mil glass or sapphire
wafers.
The chromium thickness can be varied to change the resistance per
square of the film. The thickness was changed from wafer to wafer
depending on the desired parameters and was usually between 10 and
40 nm giving a corresponding resistance of 10 to 110 $\Omega$ per
square.  The Nb meander lines are patterned using electron beam lithography
and then etched using a standard fabrication process \cite{patel2005}. The meander line has a
width of 5 $\mu m$ and a pitch of 20 $\mu m$ to obtain an
inductance of approximately $1.3$ $\mu H$. The meanders were 1 mm
wide and as long as 30 mm. In order to achieve these lengths, the 2 mm electron
beam fields were stitched together. A 250 nm
layer of $SiO_2$ is deposited on top of the niobium to protect it
from scratching and breaking which was a problem in the original
batches of filters. The filters are diced to the necessary length
and width to reach the desired response and glued to the housings
using GE varnish.

Microwaves can propagate in waveguide modes in the filter housing if
their frequency exceeds the cutoff frequency.  The cutoff frequency
for the transverse electric modes , $TE_{m,n}$, and transverse
magnetic modes, $TM_{m,n}$ is given by \cite{ramo}
\begin{equation} f_{m,n} = \frac{1}{2 \pi} \sqrt{\frac{1}{\mu \epsilon} \left[ \left(\frac{m \pi}{W}\right)^2+ \left( \frac{n \pi}{H}\right)^2 \right]} \label{cutoff} \end{equation}
where W and H are the width and height of the cavity respectively
and m and n are indices identifying the number of nodes
perpendicular to the direction of propagation, where for $TE_{m,n}$
either $m \neq 0$ or $n \neq 0$ and for $TM_{m,n}$ both $m \neq 0$
and $n \neq 0$.  In all of our filter housing designs the height of the cavity is much smaller then the width. For this geometry, Equ. \ref{cutoff} shows the cutoff
frequency for all the modes are increased by reducing the height
except the $TE_{1,0}$ which is, independent of this dimension. For
the low frequency control lines, the filter housings have dimensions $W=2.38$ mm and $H=0.30$ mm, the lowest order waveguide cutoff is $f_{1,0}=62.7$ GHz and the next lowest mode has a cutoff of $f_{1,1}=496$ GHz.  For the high frequency lines, the filter housings have dimensions $W=1.60$ mm and $H=0.46$ mm, the waveguide cutoff is $f_{1,0}=93.7$ GHz and the next lowest mode has a cutoff of $f_{1,1}=341$ GHz.

\begin{figure}
\centering
\includegraphics[width=3.0in]{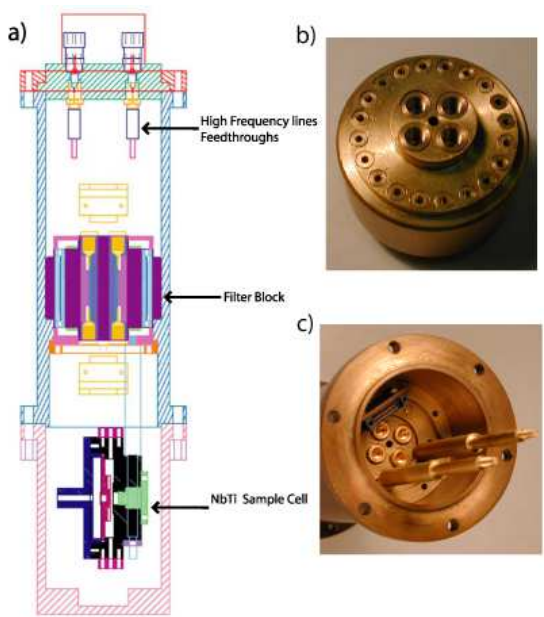}
\caption{\label{samplecan} (Color Online) a) Drawing of a cross-sectional view of the
assembled outer sample can and NbTi sample cell. b) a picture of the filter block. c) The assembled filter block within the sample vacuum can.}
\end{figure}

In order to guarantee an optimal thermalization we have housed the microstrip filters in a brass block attached to a vacuum tight sample can that can be anchored at the mixing chamber stage of a dilution refrigerator. As shown in Fig. \ref{samplecan} the sample can
consists of two chambers separated by the filter block. The filter block can house up to 20 low low frequency lines and 4 high frequency lines. The device under test is mounted in a NbTi cell that acts as a magnetic shield below its transition temperature around 10K and is located in the bottom chamber of the sample can. Coaxial cables and twisted pairs enter the can through vacuum tight epoxy feedthroughs (Stycast 2850 FT) mounted in the lid. The can is filled with $He^4$ which becomes superfluid at low temperature, providing a uniform thermalization to the microstrip filters and to the qubit chip.

\section{Testing}

\begin{figure}
\centering
\includegraphics[width=3.5in]{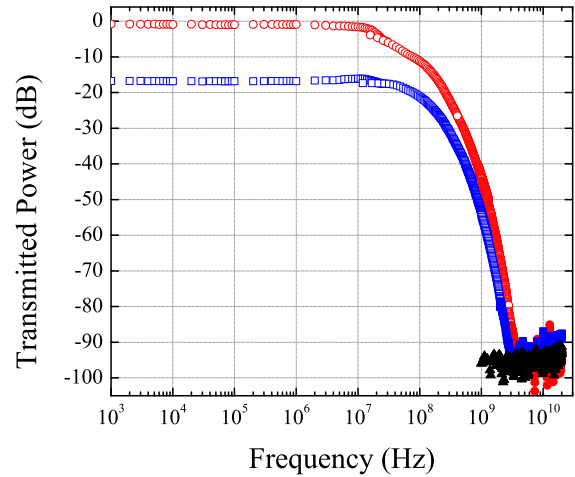}
\caption{(Color Online) The power transmitted through the lossy microstrip filter
into a 50 $\Omega$ load as a function of frequency, measured at room
temperature (blue squares)and 4.2 K (red circles)}
\label{fig:meanderfilter}
\end{figure}

We measured the filters attenuation at room and liquid helium temperature.
Figure \ref{fig:meanderfilter} shows the power transmitted through
the filter to a detector with a 50 $\Omega$ load at room temperature
(blue squares) and at 4.2 K (red circles). At high frequencies the
rf signal was generated using a HP 83731B synthesized signal
generator or a HP 8620C sweep oscillator using a HP 435A power meter
and HP 8484A diode power sensor to detect the signal. At lower
frequencies, a SRS DS345 synthesized function generator was used to
create the signal and a standard oscilloscope was used to measure
its rms amplitude. The black triangles show the background of the
power sensor. The high and low frequency data sets do not exactly
overlap since the power sensor was slightly outside its calibrated range.
\begin{figure}
\centering
\includegraphics[width=3.0in]{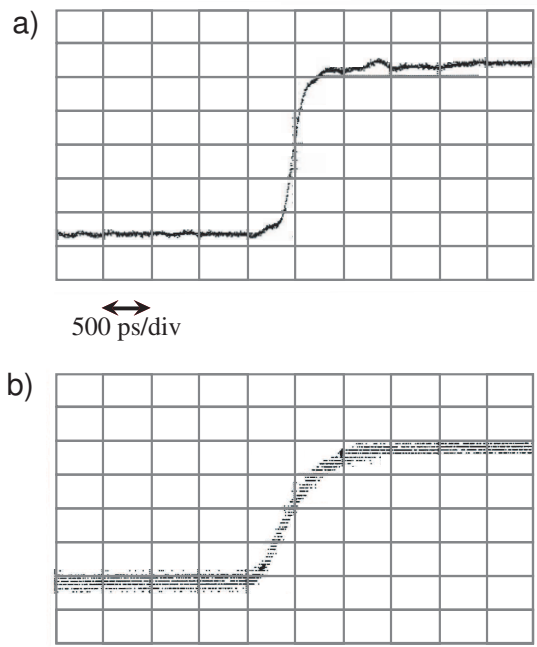}
\caption{Filter response to a DC pulse. a) input pulse with a rise time of 360ns. b) output pulse, in this case the rise time is 900ns. For both figure the horizontal scale is 500ps/div.}
\label{fig:pulse}
\end{figure}
The room temperature data shows 16 dB of attenuation below 10 MHz due to
the current primarily traveling though the resistive chromium. The
Nb meander line is not superconducting and gives a resistance larger
than the chromium. The cutoff, 71 MHz, is consistent with the
designed cutoff of the lossy microstrip line. When the Nb goes
superconducting the low frequency current is shunted through the
meander line and the attenuation is that of the cables alone. The
first cutoff at 15 MHz corresponds to the cutoff of the meander line
while the second kink is the cutoff of the microstrip. The data
shows that after the cutoff the attenuation quickly increases beyond
what can be measured using the current setup and stays below the
detection threshold to frequencies at least as high as 20 GHz. The
Nb meander lines were tested using four point measurements and are
superconducting for currents up to 10 mA.

In fig. \ref{fig:pulse} we show the response of our filters to a DC pulse. As it can be seen from the data, the filters modify the rise time of the pulse from 360ps to 900ps. This is an encouraging result since to test the quantum behavior of a superconductive qubit it is desirable to be able to send pulses with a fast rise time of less than 1ns \cite{bennett2007}.

\section{Conclusion}

The criteria for filters to isolate a qubit from the higher temperature dissipative environment are quite stringent, especially when coupled with the need to get high frequency signals to the qubit for control and readout.  We have designed a special type of filter, a lossy transmission line, which cuts off very sharply at the design frequency and has no passbands above the cutof. The lossy microstrip design provides a flexible way to effectively filter many signals over a wide ranges of frequencies. The cutoff and attenuation level of the filters can be controlled by design using the dimensions and the resistivity of the microstrip. These filters are easy to fabricate and install making them a useful tool for measurements of quantum behavior of superconducting devices.\\

\begin{acknowledgments}
This work was supported in part by NSF and by AFOSR and NSA through
a DURINT program.
\end{acknowledgments}

\end{document}